\documentclass[12pt]{article}
\begin{document}

\title{On recent data of branching ratios for \\
$\psi(2S)\rightarrow J/\psi\pi\pi$ and $\psi(2S)\rightarrow J/\psi\eta$}

\author{Y. ~F. ~Gu$^{a}$ \thanks{Email: guyf@bepc5.ihep.ac.cn}, 
\hspace{0.3cm}    X. ~H. ~Li$^{b}$ 
\thanks{Email: lixh@sseos.lbl.gov. On leave from IHEP, Beijing.}}

\date{}
\maketitle


\begin{center}
{\small   $^a$Institute of High Energy Physics,
  Beijing 100039, People's Republic of China  

$^b$Nuclear Science Division, Lawrence Berkeley National Laboratory, \\
 University of California, Berkeley, CA 94720, USA  }
\end{center}

\vspace{0.5cm}
\begin{center} Dec. 23, 1998  \end{center}
\vspace{1.5cm}

\begin{abstract}
Recent data on branching ratios for $\psi(2S)$ decays to $J/\psi\pi^+\pi^-$, 
$J/\psi\pi^0\pi^0$ and $J/\psi\eta$ are reviewed.  An alternative treatment 
of data is proposed to get rid of the logical inconsistency which occurs in 
original computational procedure. 
\end{abstract}

\newpage

There is a group of exclusive $\psi(2S)$ decays into $J/\psi$ whose 
branching ratios are of importance to both charmonium and other (e.g. B and 
relativistic nuclear collision) physics, 
but were mostly measured two decades ago [1].  Recently, 
new results were reported on the branching ratios of 
$\psi(2S)\rightarrow J/\psi\pi^+\pi^-$, $\psi(2S)\rightarrow J/\psi\pi^0\pi^0$,
and $\psi(2S)\rightarrow J/\psi\eta$ by the E760 experiment [2], and on the 
ratio of $B(\psi(2S)\rightarrow J/\psi\pi^+\pi^-)$/$B(\psi(2S)\rightarrow
\mu^+\mu^-)$ by the E672/E706 experiment [3].  With these new entries, 9 
branching ratios of the $\psi(2S)$ are reanalyzed by the Particle Data Group
(PDG) [1]. In this letter, we comment on these recent data and propose, among
other things, an alternative way of treating the data to get rid of the 
logical inconsistency which occurs in 
the computational procedure of ref. [2].
 
In determining the branching ratios of $\psi(2S)\rightarrow f$, where f is
$J/\psi\pi^+\pi^-$, $J/\psi\pi^0\pi^0$, or $J/\psi\eta$, the authors of 
ref. [2] used the expression 
\begin{equation}
B(\psi(2S)\rightarrow f)= \frac{\varepsilon_{J/\psi X}}{\varepsilon_f}
\frac{N_f}{N_{J/\psi X}} B(\psi(2S)\rightarrow J/\psi X), 
\end{equation}
where $\varepsilon_{f}$ and $\varepsilon_{J/\psi X}$
include the geometrical acceptance and efficiencies for triggering and
selection of the exclusive (f with $\pi^0$ or $\eta\rightarrow\gamma\gamma$) 
and inclusive $(J/\psi X$ with $J/\psi\rightarrow e^+e^-)$ decays of the
$\psi(2S)$, respectively; $N_{f}$ and $N_{J/\psi X}$ are the numbers of 
exclusive and inclusive events selected, respectively. The authors claimed 
that, by determining 
$\frac{\varepsilon_{J/\psi X}}{\varepsilon_{f}}\frac{N_{f}}{N_{J/\psi X}} $
and using the branching ratio of $\psi(2S)\rightarrow J/\psi X$ in the PDG, 
they are able to
make measurements of $B(\psi(2S)\rightarrow J/\psi\pi^+\pi^-)$ and
$B(\psi(2S)\rightarrow J/\psi\pi^0\pi^0)$ with errors comparable to the world
average.  However, one notes that the method used by the authors to 
treat their data is questionable.

 As a matter of fact, the PDG value of $B(\psi(2S)\rightarrow J/\psi X)$,
$0.57\pm 0.04$, on which the authors rely for determining the branching
ratios of $B(\psi(2S)\rightarrow J/\psi\pi^+\pi^-)$,
$B(\psi(2S)\rightarrow J/\psi\pi^0\pi^0)$, and
$B(\psi(2S)\rightarrow J/\psi\eta)$, is not an independent, direct measurement,
but is a value from a constrained fit to 7 branching ratios for the $\psi(2S)$ 
of 13 significant measurements including 
$B(\psi(2S)\rightarrow J/\psi\pi^+\pi^-)$,
$B(\psi(2S)\rightarrow J/\psi\pi^0\pi^0)$ and
$B(\psi(2S)\rightarrow J/\psi\eta)$ [4], 
which are what the authors attempt to
measure. One finds here an apparent logical inconsistency. 

 A correct way to make the measurements self-consistent  
would be to solve a linear equation of the form 
\begin{equation}
 x=a(x+b),
\end{equation}
where 
$x\equiv B(\psi(2S)\rightarrow J/\psi\pi^+\pi^-)+B(\psi(2S)\rightarrow J/\psi
\pi^0\pi^0)+B(\psi(2S)\rightarrow J/\psi\eta)$, 
$a \equiv \sum\limits_{i=1}^{3} 
\frac{\varepsilon_{J/\psi X}}{\varepsilon_{fi}}
\frac{N_{fi}}{N_{J/\psi X}}$,  and $b \equiv
0.273 B(\psi(2S)\rightarrow \gamma\chi_{c1}) +
0.135 B(\psi(2S)\rightarrow \gamma\chi_{c2})$
using the PDG data which are irrelevant to $x$.
The solutions of Eq. (2) for $x$ with the addition of b
give $ B(\psi(2S)\rightarrow J/\psi X)=0.315\pm0.226$  
and $0.292\pm0.136$ for the two data sets (1990 and 1991) of the E760 
experiment, respectively, which can now be applied to Eq. (1) consistently. 
Table 1 gives the computed branching ratios for three
exclusive decays. The same results can be achieved by using 
an iterative method of solving Eq. (2).  
However, the results thus obtained are much worse than the PDG world 
averages or fit values.  Here the substantial errors are mainly due to the 
enlargement of the uncertainty  by a factor of $\frac{1}{1-a}\approx 9$. 
Such an approach has practically nothing to recommend it. 

We would thus propose that, as a natural and probably the best way to deal 
with the E760 data, its measured ratios
$\frac{\varepsilon_{J/\psi X}}{\varepsilon_{f}}\frac{N_{f}}{N_{J/\psi X}}$
be used directly as the ratios of two branching ratios,
$\Gamma(\psi(2S)\rightarrow f)/ \Gamma(\psi(2S)\rightarrow J/\psi X)$
$(f=\pi^+\pi^-$, $\pi^0\pi^0$ or $\eta)$. 
With these new entries, in addition to the previous 13 measurements [4],
an overall fit as used by the PDG is redone with minor
simplification [5]. 
In order to make the input data of E760 independent, two uncorrelated ratios 
presented in Table 2, 
 \begin{equation}
 \Gamma(\psi(2S)\rightarrow J/\psi\pi^0\pi^0)/
  \Gamma(\psi(2S)\rightarrow J/\psi\pi^+\pi^-) =
  \frac{\varepsilon_{J/\psi \pi^+\pi^-}}{\varepsilon_{J/\psi \pi^0\pi^0}}
  \frac{N_{J/\psi \pi^0\pi^0}}{N_{J/\psi \pi^+\pi^-}} 
\end{equation} 
 and
 \begin{equation}
 \Gamma(\psi(2S)\rightarrow J/\psi\eta)/
  \Gamma(\psi(2S)\rightarrow J/\psi X) =
  \frac{\varepsilon_{J/\psi X}}{\varepsilon_{J/\psi \eta}}
  \frac{N_{J/\psi \eta}}{N_{J/\psi X}}, 
\end{equation}
are in fact used in the fit instead of three ratios of
$\Gamma(\psi(2S)\rightarrow J/\psi$ $f)/\Gamma(\psi(2S)\rightarrow J/\psi X) $
($f=\pi^+\pi^-$, $\pi^0\pi^0$ and $\eta$). 
Table 3 summarizes the fit results. The 1996 PDG fit values [4] and the E760
reporting values [1] are also included for comparison. 
Using the combined fit values 
in Table 3, along with the 1998 PDG averages for
$B(\chi_{c1}\rightarrow\gamma J/\psi)$, $B(\chi_{c2}\rightarrow\gamma J/\psi)$,
$B(\pi^{0}\rightarrow\gamma \gamma)$, $B(\eta\rightarrow neutral$ $modes)$ [6],
and $B(J/\psi\rightarrow e^+e^-)$ [7], 
one may also compute $B(\psi(2S)\rightarrow  J/\psi X)$ to be $0.57\pm 0.04 $,
$B(\psi(2S)\rightarrow J/\psi$ $neutrals)$ to be $0.235\pm 0.026 $,
and $B(\psi(2S)\rightarrow e^+e^-)$ to be $(8.4\pm0.8)\times 10^{-3}$.  

We do not recommend using the 1998 PDG fit values of branching ratios for the
$\psi(2S)$ decays to $J/\psi$ and anything [1].  
The ratio of the two $\psi(2S)$ decay partial widths, 
$B(\psi(2S)\rightarrow J/\psi\pi^+\pi^-)/
 B(\psi(2S)\rightarrow \mu^+\mu^-)$, 
measured by the E672/E706 experiment [3] was mistaken for 
$B(\psi(2S)\rightarrow J/\psi\pi^+\pi^-)
/B(\psi(2S)\rightarrow J/\psi\mu^+\mu^-)$ in ref. [1].  In addition, the
E760 data in their present form are inappropriate for such a fit by the
above-mentioned argument.


\bigskip
The authors wish to thank their colleagues on the BES collaboration 
for useful discussions. This work was supported by the 
National Natural Science Foundation of 
China under Contract No. 19290400, the Chinese Academy of Sciences under
Contract No. K10,  the U.S. Department of Energy 
under Contract No. DE-AC03-76SF00098.

\clearpage

\begin{table}
\caption{$\psi(2S)$ branching ratios determined by solving a 
linear equation of the form (2).}
\begin{center}
\begin{tabular}{cccc} \\ \hline
Channel  & 1990  & 1991  \\ \hline 
$\psi(2S)\rightarrow J/\psi\pi^+\pi^-$&$0.136\pm0.099$ & $0.156\pm0.074$ \\
$\psi(2S)\rightarrow J/\psi\pi^0\pi^0$&$0.118\pm0.087$ & $0.084\pm0.041$ \\
$\psi(2S)\rightarrow J/\psi\eta$& $0.025\pm0.021$ & $0.017\pm0.009$ \\ \hline
\end{tabular}
\end{center}
\end{table}

\begin{table}
\caption{Ratios from the E760 measurement as input data of fit.}

\begin{tabular}{ccccc} \\ \hline
Ratio  & 1990  & 1991 & Combined \\ \hline
$\Gamma(\psi(2S)\rightarrow J/\psi\pi^0\pi^0)/
 \Gamma(\psi(2S)\rightarrow J/\psi\pi^+\pi^-)$ & 
$0.868\pm0.171$ & $0.539\pm0.089$ & $0.609\pm 0.079$  \\
$\Gamma(\psi(2S)\rightarrow J/\psi\eta)/
 \Gamma(\psi(2S)\rightarrow J/\psi X)$  &
$0.080\pm0.033$ & $0.056\pm0.018$ & $0.062\pm0.016$ \\ \hline
\end{tabular}
\end{table}

\begin{table}
\caption{$\psi(2S)$ branching ratios by a simutaneous least-square fit. }
{\small
\begin{tabular}{cccccc} \\  \hline
Channel  & 1990  & 1991 & Combined & PDG96 & E760 \\ \hline
\ $\psi(2S)\rightarrow J/\psi\pi^+\pi^-$  & $0.318\pm0.025$ & $0.323\pm0.025$
                     & $0.318\pm0.025$ & $0.324\pm0.026$ & $0.283\pm0.029$ \\
\ $\psi(2S)\rightarrow J/\psi\pi^0\pi^0$  & $0.196\pm0.025$ & $0.178\pm0.022$
                     & $0.186\pm0.021$ & $0.184\pm0.027$ & $0.184\pm0.023$ \\
\ $\psi(2S)\rightarrow J/\psi\eta$        & $0.027\pm0.003$ & $0.027\pm0.003$
                     & $0.027\pm0.003$ & $0.027\pm0.004$ & $0.032\pm0.010$ \\
\ $\psi(2S)\rightarrow \gamma\chi_{c1}$   & $0.086\pm0.008$ & $0.087\pm0.008$
                     & $0.087\pm0.008$ & $0.087\pm0.008$ &  \\
\ $\psi(2S)\rightarrow \gamma\chi_{c2}$   & $0.078\pm0.008$ & $0.078\pm0.008$
                     & $0.078\pm0.008$ & $0.078\pm0.008$ &  \\ \hline
\end{tabular}}
\end{table}


\begin{thebibliography}{99}
\bibitem{5} We ignore the one constraint and fit to determine 5 parameters
without the ``other fit modes''. A check using 13 measurements contained in
[4] shows no essential difference in fit quality,  
results and errors between the PDG fit
and our simplification, except for $B(\psi(2S)\rightarrow J/\psi\eta)=
0.027\pm0.003$ obtained by us compared with $0.027\pm0.004$ in ref. [4].
\bibitem{6} We use the new fit value for
$B(\eta\rightarrow neutral$ $modes)$ given in ref. [1],  $0.715\pm0.006$,
instead of the previous fit value, $0.708\pm 0.008$, which is still used
by the PDG in calculating $B(\psi(2S)\rightarrow J/\psi$ $neutral$ $modes)$ 
in ref. [4].
\bibitem{7} We use the new fit value for
$B(J/\psi\rightarrow e^+e^-)$ given in ref. [1] or [4],  
$(6.02\pm0.19)\times 10^{-2} $,
instead of the previous fit value,
$(5.99\pm0.25)\times 10^{-2} $, which is still used
by the E760 in calculating $B(\psi(2S) \rightarrow e^+e^-)$ in ref. [2].
\end{thebibliography}
\end{document}